\documentclass{article}
\usepackage{color}
 \usepackage{amsbsy}
  \usepackage{bm}
 \usepackage{amsfonts}
\usepackage{amssymb}
\usepackage{amsmath}
\usepackage{graphics}
\usepackage{graphicx}
\begin{document}

\title{Modification of the masses of the lightest neutral mesons in a hadronic medium under an external magnetic field }
\author{R. M. Aguirre}

\date{
\it{Departamento de Fisica, Facultad de Ciencias Exactas}, \\\it{Universidad Nacional de La Plata,} \\
\it{and IFLP, UNLP-CONICET, C.C. 67 (1900) La Plata, Argentina.}}


\maketitle

\begin{abstract}
The effective masses of the neutral mesons in a hadronic medium
and under an external magnetic field are evaluated as functions of
the baryonic density and the field intensity. For this purpose the
meson polarization is evaluated in the one loop approximation
using a Quantum Hadrodynamics model which includes $\pi, \sigma,
\omega$, and $\rho$ mesons. The propagators of the baryons include
the full effect of the coupling to the magnetic field through
their charges and their anomalous magnetic moments. Within the
range of magnetic intensities considered here $10^{17}$ G $< B <
10^{19}$ G, the dependence on $B$ is moderate for the pion and the
longitudinal component of the $\omega$ meson, and negligible for
the remaining mesons.

\noindent
\\

\end{abstract}

\newpage

\section{Introduction}

The physics of matter subject to strong magnetic fields presents
multiple aspects, which is why it has been widely studied in the
past \cite{LAI,MIRANSKY}, and new facets are always being
investigated.

Among the different empirical manifestations of very strong
magnetic fields, two situations have received particular attention
in recent years. \\
On one hand the presence of intense magnetic fields has been
deduced from the observational data of certain compact stars which
have been generally included within the magnetar model
\cite{DUNCAN,THOMPSON}. The sustained X-ray luminosity in the soft
(0.5-10 keV) or hard (50-200 keV) spectrum, as well as the
bursting activity of these objects have been attributed to the
dissipation and decay of very strong fields. The intensity of
these fields has been estimated around $10^{15}$ G at the star
surface, but could reach much higher values in the dense interior
of the star.  The origin of the magnetism, however, is still under
debate \cite{RUDERMAN,THOMPSON2,VIDANA}.\\
Furthermore, extreme magnetic intensities are expected in heavy
ion collisions \cite{KHARZEEV,MO,SKOKOV}. Experimental evidence of
this fact is the preferential emission of charged particles along
the direction of the magnetic field for noncentral heavy ion
collisions, due to magnetic intensities $e\, B \sim 10^2$
MeV$^2$\cite{KHARZEEV}. In spite of its evanescent character, the
strong field could influence the hadronization process
\cite{FUKUSHIMA}.\\

The effect of external magnetic fields on the quark structure of
mesons has intensively been studied in recent years by using
lattice QCD \cite{LatticeQCD}, QCD sum rules for heavy mesons
\cite{QCDsumrules}, non-relativistic potential for heavy mesons
\cite{YOSHIDA}, an effective hamiltonian with QCD basis
\cite{ANDREICHIKOV}, and the Nambu and Jona-Lasinio model
\cite{NJL1,NJL2}. Some of these descriptions do not consider quark
confinement explicitly, and others lack of the interaction among
hadrons.\\
On the other hand, the use of effective hadronic models to study
pions \cite{ANDERSEN,COLUCCI,CHEOUN,ADHYA,AGUIRRE,MANDAL} as well
as vector mesons \cite{KAWAGUCHI,GHOSH}, takes account of the
hadronic environment although it does not include the substructure dynamics.\\
It is reasonable that light mesons experience more noticeably the
effects of an external magnetic field if $e B \sim m_\pi^2$ ($B
\sim 3 \times 10^{18}$ G). Furthermore, as the neutral mesons do
not couple directly with the magnetic field, any change in their
dynamical properties is due to their interactions with the
hadronic medium.\\
The aim of this work is to investigate the effect of a uniform
external field on the properties of the lightest neutral meson. In
our approach the magnetic field is treated as a classical external
field, therefore we neglect electromagnetic quantum corrections.
Otherwise, the mixing of neutral pions and photons could be a
source of level repulsion.\\
We consider the magnetic intensity, the baryonic number density,
and
the isospin composition of matter as relevant parameters.\\
A successful description of the dense hadronic environment has
been given by  a covariant model of the hadronic interaction known
as Quantum Hadro-Dynamics (QHD) \cite{SW}. It has been used to
study the structure of neutron stars and particularly to analyze
hadronic matter in the presence of an external magnetic field
\cite{CHAKRABARTY,BRODERICK,MALLICK,DONG1,DONG2,AB&V,ISAAC,REZAEI,CENTELLES}.
In this formulation the elementary degrees of freedom are hadronic
fields, regarded as structureless particles.  For this reason when
the quark dynamics is expected to manifest, additional input
should be included in the model. However, QHD is an adequate tool
to analyze many-body effects in the hadronic
medium for a wide range of applications.\\
The versatility of QHD allows the inclusion of the intrinsic
magnetic moments in a covariant way. Due to the strength of the
baryon-meson couplings, the mean field approximation (MFA) is
usually employed. Within this approach the meson fields are
replaced by their expectation values and assimilated to a
quasi-particle picture of the baryons. Finally the meson mean
values are obtained by solving the classical meson equations
taking as sources the baryonic currents. This scheme is
conceptually clear and easy to implement.\\
We use here a model including pions, $\sigma, \omega$ and $\rho$
mesons in order to evaluate the meson polarization in the one loop
approximation. The effective mass of the mesons in the hadronic
medium is defined and analyzed at zero temperature for a wide
range of densities $0 < n_B < 3 n_0$, with $n_0$ the saturation
density of nuclear matter, and magnetic intensities $10^{16}$ G
$\leq B \leq 10^{19}$ G. Two different isospin composition of
matter are considered, isospin symmetric nuclear matter and
neutral nuclear matter in equilibrium against beta decay.\\
As mentioned before, the model is not able to describe the
consequences of strong magnetic fields on the quark structure.
Hence we will not consider effects like the mixing of
quark-antiquark bound states induced by an external field at zero
baryonic density as reported in recent works \cite{HEAVY}.

The main contribution to our calculations comes from a ring
diagram of the nucleon propagator, which includes the full effect
of the coupling of the external field through the electric charge
and the
anomalous magnetic moment.\\
The significative role played by the intrinsic magnetic moments of
the hadrons has been pointed out for the evaluation of bulk
properties of dense nuclear matter under strong
magnetic fields \cite{DONG1,DONG2,AB&V,ISAAC,REZAEI}.\\
The propagator of a proton with anomalous magnetic moment immersed
in a uniform magnetic field has been presented in \cite{CHEOUN},
and extended to  finite density and temperature  in
\cite{AGUIRRE}.

This work is organized as follows. In the next section the
one-loop polarization insertion is presented. The results and
discussion are given in Sec. III, and the conclusions are shown in
Sec. IV.

\section{In-medium meson polarization insertion}\label{Sec1}

The effective model for the hadronic interaction is given by
\begin{eqnarray}
\mathcal{L}&=&\sum_{a=n,p}
\bar{\Psi}^a\Big[\gamma_\mu\left(i\,\partial^\mu-q_a\,
A^\mu+g_\omega \omega^\mu+\frac{g_\rho}{2} \bm{\tau} \cdot
\bm{\rho}^\mu-\frac{g_A}{2 f_\pi} \gamma_5 \bm{\tau} \cdot
\partial^\mu \bm{\phi}-\frac{1}{4 f_\pi^2}\bm{\tau} \cdot \bm{\phi} \times \partial^\mu \bm{\phi}\right)
\nonumber \\ &-& m_0+g_\sigma
\sigma+\frac{\kappa_a}{2}\,\sigma^{\mu\nu}\,\mathcal{F}_{\mu\nu}\Big]\Psi^a
-\frac{A}{3}\, \sigma^3-\frac{B}{4} \,\sigma^4+\frac{C}{4} \,
(\omega_\mu \omega^\mu)^2+ D \bm{\rho}_\lambda \cdot
\bm{\rho}^\lambda \; \omega_\mu \omega^\mu +\mathcal{L}_M
\nonumber 
\end{eqnarray}

here $\mathcal{L}_M$ stands for the free mesons part, and only the
lowest lying baryons are considered with anomalous magnetic moment
represented by $\kappa_a$. The interaction includes one and two
pion vertices, and the self-interaction of the $\sigma$ and
$\omega$ mesons, together with the $\omega - \rho$ coupling
\cite{HOROWITZ,PIEKAREWICZ}.

In our approach, the fundamental state of matter is given by a
mean field approach (MFA), which is equivalent to include the
tadpole diagram (see Fig.1a) in a self-consistent solution but
neglecting divergent contributions coming from the Dirac sea. At
this step it is assumed that meson propagation is not modified by
the hadronic interaction. The effect of the magnetic field,
instead, is fully included for both meson and nucleon propagators.
It can be verified that pions do not contribute to the tadpole
diagram, since the pion-nucleon vertices depend on the transferred
pion momentum. Furthermore the neutral mesons $\pi^0,\, \sigma, \,
\omega_\mu$ and $\rho_\mu^0$ are not affected directly by the
magnetic field.\\
At this step a quasi-particle picture is obtained for the
nucleons, with effective mass $m=m_0-g_\sigma S$, and energy
spectra $p_0^{(a)}=g_\omega W+g_\rho R I_a\pm E_a$, with
\begin{eqnarray}
E_1&=&\sqrt{p_z^2+(\Delta_n-s\,\kappa_1 B)^2}
\nonumber \\
\Delta_n&=&\sqrt{m^2+2 n q B} \nonumber
\end{eqnarray}
$I_1=1$ for protons, and
\begin{eqnarray}
E_2&=&\sqrt{p_z^2+(\Delta-s\,\kappa_2 B)^2}\label{Nspectra} \nonumber \\
\Delta&=&\sqrt{m^2+p^2_x+p^2_y} \nonumber
\end{eqnarray}
$I_2=-1$, for neutrons. The index $s=\pm 1$ indicates spin
projection along the direction of the uniform magnetic field, and
the discrete index $n$ for protons, comes from the Landau
quantization. It must be remembered that the lowest Landau level
$n=0$ admits only $s=1$.\\
Furthermore, the quantities $S, \, W$, and $R$ correspond
respectively to the in-medium expectation values of the $\sigma$
and time-like components of $\omega$ and $\rho^0$ mesons
\cite{SW}. They are related to the hadronic densities by
\begin{eqnarray}
\left(m_\sigma^2+ A S + B S^2 \right) S&=&g_\sigma (n_{s1}+n_{s2})
\nonumber \\
\left(m_\omega^2+ C W^2 + 2 D R^2 \right) W&=&g_\omega (n_1+n_2)
\nonumber \\
\left(m_\rho^2+  2 D W^2 \right) R&=&g_\omega (n_1-n_2) \nonumber
\end{eqnarray}
 where
\begin{eqnarray}
n_1&=&\frac{q B}{2 \pi^2}\sum_{n,s}  \int dp_z \left[n_F(E_1,\mu_1)-n_F(-E_1,\mu_1)\right]\nonumber \\
n_2&=&\sum_s \int \frac{d^3p}{(2 \pi)^3}
 \left[ n_F(E_2,\mu_2)-n_F(-E_2.\mu_2)\right]\nonumber \\
n_{s1}&=&\frac{q B}{2 \pi^2} \, m\, \sum_{n, s} \int dp_z
\frac{\Delta_n+s\,\kappa_1
B}{E_1\,\Delta_n}[n_F(E_1,\mu_1)+n_F(-E_1,\mu_1)] \nonumber \\
n_{s2}&=&\sum_s\int \frac{d^3p}{(2 \pi)^3}
 \frac{\Delta+s\,\kappa_2 B}{E_2\,\Delta}\left[n_F(E_2,\mu_2)+n_F(-E_2,\mu_2)\right]\nonumber
\end{eqnarray}
The two first equations relate the conserved baryon number with
the chemical potentials $\mu_a$.

The nucleon propagators corresponding to this approach are given
in \cite{AGUIRRE}. For the sake of completeness we show here the
neutron propagator
\begin{eqnarray}
G^{(2)}(x',x)= \sum_s \int \frac{d^4p}{(2 \pi)^4} e^{-i
p^\mu\,(x_\mu '-x_\mu)}
\Lambda_s\left[\frac{1}{p_0^2-E_2^2+i\epsilon}+ 2
\pi\,i\,n_F(p_0)\,\delta(p_0^2-E_2^2)\right] \nonumber
\end{eqnarray}
 where
\begin{eqnarray}
\Lambda_s=\frac{ s}{2 \Delta}i\; \gamma^1 \gamma^2\left[ \not \!
u+ i \gamma^1 \gamma^2 (s \Delta-\kappa_2 B)\right] \left( \not \!
v+m+ i s \Delta \gamma^1 \gamma^2\right)\nonumber 
\end{eqnarray}
and the proton propagator
\begin{equation}
G^{(1)}(x',x)=e^{i \Phi } \int \frac{d^4 p}{(2 \pi)^4} e^{-i
p^\mu\,(x'_\mu-x_\mu)} \left[ G_0(p)+\sum_{n,s} G_{n s} (p)\right]
\label{DefGP0}
\end{equation}
where the phase factor $\Phi=q B(x+x')(y'-y)/2$ embodies the gauge
fixing. We have separated the lowest Landau level contribution
\begin{eqnarray}
G_0(p)&=&2 e^{-p_\bot^2/q B}\Lambda_0
\left[\frac{1}{p_0^2-E_0^2+i\epsilon}+ 2
\pi\,i\,n_F(p_0)\,\delta(p_0^2-E_0^2)\right] \nonumber
\\
\Lambda_0&=&\left( \not \! u+m-\kappa_1\,B\right) \Pi^{(+)}
\nonumber
\end{eqnarray}
from the higher Landau levels contributions
\begin{eqnarray}
G_{n s}(p)&=& e^{-p_\bot^2/q B}\,
\Lambda_{ns}\,\left[\frac{1}{p_0^2-E_{n s}^2+i\epsilon}+2
\pi\,i\,n_F(p_0)\,\delta(p_0^2-E_{n s}^2)\right] \nonumber \\
\Lambda_{ns}&=& (-1)^n\frac{\Delta_n+s m}{ \Delta_n}\Big\{( \not
\! u-\kappa_1 B+s \Delta_n) \Pi^{(+)} L_n(2 p_\bot^2/q B)
\nonumber\\
&-& ( \not \! u+\kappa_1 B-s \Delta_n) \Pi^{(-)} \frac{s
\Delta_n-m}{s \Delta_n+m} L_{n-1}(2 p_\bot^2/q B)\nonumber \\
&+& \left[ \not \! u+i \gamma_1 \gamma_2 (s \Delta_n-\kappa_1
B)\right] i \gamma^1 \gamma^2 \not \! v \,\frac{s \Delta_n-
m}{2\,p_\bot^2} \left[ L_n(2 p_\bot^2/q B)-L_{n-1}(2 p_\bot^2/q
B)\right]\Big\} \nonumber 
\end{eqnarray}
here $L_m$ stands for the Laguerre polynomial of order $m$, and
$p_\bot^2=p_x^2+p_y^2$.\\
For this set of equations the notation $\not \!\!
u=p_0\gamma^0-p_z\gamma^3$, $\not \! \!v=-p_x \, \gamma^1-p_y\,
\gamma^2$ $\Pi^{(\pm)}=(1\pm i\gamma^1\gamma^2)/2$ is introduced.

The results for these propagators combine the gauge invariance of
the proper time method \cite{SCHWINGER} with the momentum
representation of \cite{CHODOS} and furthermore include the
contributions of the anomalous magnetic moments. The extension to
finite densities and temperatures \cite{A&D2016} has been made in
the context of the real time formalism of Thermo-Field Dynamics,
however we only exhibit the (1,1) component which suffices for the
present calculations at zero temperature.

The magnetic field has a direct coupling to the charged mesons,
and there are also quantum corrections which, for the model
proposed and at the one loop level are described by the diagrams
of Figs. 1b-1f. The cases (d-f) are first order and comes from the
two-pion coupling to nucleons (d), the self-interaction of fourth
order, either $\sigma$ or $\omega$ (e), and the $\omega - \rho$
coupling (f). The two last cases gives zero contribution after
regularization. On the other hand, the diagrams (b) and (c) are
second order. The third order $\sigma$ self-interaction gives rise
to (c), it is divergent and needs to be regularized. Finally the
nucleon ring diagram (b), as well as the case (d) contain
divergent contributions coming from the Dirac sea which will be
omitted in our approach.\\
The finite contribution of diagram (c) is obtained by dimensional
regularization and substraction at the point
$\textsf{p}^2=m_\sigma^2$, it is given by
\begin{eqnarray}
\Pi_\sigma(\textsf{p})=\left(\frac{A}{12 \pi}\right)^2\left[
\sqrt{\frac{4 m_\sigma^2}{\textsf{p}^2}-1} \; \arctan\left(\frac{4
m_\sigma^2}{\textsf{p}^2}-1\right)^{-1/2}-\frac{\pi}{2
\sqrt{3}}\right]  \label{3Sigma}
\end{eqnarray}
within the regime $0 < \textsf{p}^2 < 4 m_\sigma^2$.

 The correction (d) to the pion propagation has
been evaluated in Ref. \cite{AGUIRRE} and gives zero contribution
for the neutral pion.

Finally we consider the nucleon ring diagram (b). As we are
interested in corrections to the neutral mesons, there is no
neutron-proton mixing at the vertices. \\
The polarization insertion can be classified as direct
\begin{eqnarray}
i \Pi_\pi(p)&=&\left( \frac{g_A}{2 f_\pi}\right)^2 p_\mu p_\nu
\sum_{a=1,2} \int \frac{d^4q}{(2 \pi)^4} Tr\left\{\gamma^\mu
\gamma_5 G^{(a)}(q) \gamma^\nu \gamma_5 G^{(a)}(q-p)\right\}
\nonumber \\
%
i \Pi_\sigma(p)&=&g_s^2 \sum_{a=1,2}  \int \frac{d^4q}{(2 \pi)^4}
Tr\left\{ G^{(a)}(q)  G^{(a)}(q-p)\right\} \label{PolSigma} \\
i \Pi_\omega^{\mu \nu}(p)&=&g_w^2 \sum_{a=1,2}  \int
\frac{d^4q}{(2 \pi)^4} Tr\left\{ \gamma^\mu G^{(a)}(q)\gamma^\nu
G^{(a)}(q-p)\right\} \label{PolOPE} \\
i \Pi_\rho^{\mu \nu}(p)&=&\frac{g_r^2}{4} \sum_{a=1,2}  \int
\frac{d^4q}{(2 \pi)^4} Tr\left\{ \gamma^\mu G^{(a)}(q)\gamma^\nu
G^{(a)}(q-p)\right\} \label{PolOPE}
\end{eqnarray}
which describes the propagation of a given class of mesons, and
the mixing components
\begin{eqnarray}
i \Pi_{\omega \rho}^{\mu \nu}(p)&=&g_w \frac{g_r}{2} \sum_{a=1,2}
I_a \int \frac{d^4q}{(2 \pi)^4} Tr\left\{ \gamma^\mu
G^{(a)}(q)\gamma^\nu G^{(a)}(q-p)\right\} \label{PolMixVecT} \\
i \Pi_{\omega \rho}(p)&=&g_w \frac{g_r}{2} \sum_{a=1,2} I_a \int
\frac{d^4q}{(2 \pi)^4} Tr\left\{ \gamma^3
G^{(a)}(q)\gamma^3 G^{(a)}(q-p)\right\} \label{PolMixVecL} \\
i \Pi_{\pi \omega}(p)&=& g_w \frac{g_A}{2 f_\pi} p_\mu
\sum_{a=1,2} \int \frac{d^4q}{(2 \pi)^4} Tr\left\{\gamma^\mu
\gamma_5 G^{(a)}(q) \gamma^3 G^{(a)}(q-p)\right\} \label{PolPiOme} \\
i \Pi_{\pi \rho}(p)&=& g_r \frac{g_A}{4 f_\pi} p_\mu \sum_{a=1,2}
I_a \int \frac{d^4q}{(2 \pi)^4} Tr\left\{\gamma^\mu \gamma_5
G^{(a)}(q) \gamma^3 G^{(a)}(q-p)\right\} \label{PolPiRho}
\end{eqnarray}
which describes the conversion between different classes of
mesons. For the transversal vector mixing of Eq.(\ref{PolMixVecT})
only $\mu, \, \nu=1,2$ is possible. Explicit expressions are given
in the Appendix.\\
There are other mixing components not enumerated here, but we
focus only on those which are useful for the present calculations.
As the meson effective masses will be defined for the dynamical
regime ${\bm p}=0$, we have found that only the mixing components
shown above give non-zero contribution.\\
Thus, there are three blocks of generalized polarization at ${\bm
p}=0$.  The sigma component alone on one hand, the transversal
$\omega$ and $\rho$ components together with the transversal mix
of Eq.(\ref{PolMixVecT}) on the other hand, and finally the pion,
with the (3,3) component of the $\omega$ and $\rho$ together with
the vector longitudinal mixing of Eq.(\ref{PolMixVecL}) and the
pion-vector mixing of Eqs.(\ref{PolPiOme}) and (\ref{PolPiRho}).\\
For each of these blocks we define a dielectric function
\begin{eqnarray}
\varepsilon_S(p)&=&\textsf{p}^2-m_\sigma^2-\Pi_\sigma(p) \label{DielS}\\
\varepsilon_T(p)&=&\det M_T  \label{DielT}\\
\varepsilon_L(p)&=&\det M_L  \label{DielL}
\end{eqnarray}

where $\Pi_\sigma$ consists of the sum of Eqs.(\ref{3Sigma}) and
(\ref{PolSigma}), and
\begin{eqnarray}
M_T&=& \left(
\begin{array}{cccc}
\textsf{p}^2-m_\omega^2-\Pi_\omega^{1\,1}&-\Pi_\omega^{1
\,2}&-\Pi_{\omega \rho}^{1 \,1}&-\Pi_{\omega \rho}^{1 \,2}\\ \\
-\Pi_\omega^{2\,1}&\textsf{p}^2-m_\omega^2-\Pi_\omega^{2\,2}&-\Pi_{\omega
\rho}^{2 \,1}&-\Pi_{\omega \rho}^{2 \,2}\\ \\
-\Pi_{\omega \rho}^{1\,1}&-\Pi_{\omega \rho}^{2
\,1}&\textsf{p}^2-m_\rho^2-\Pi_\rho^{1\,1}&-\Pi_\rho^{1\,2}\\ \\
-\Pi_{\omega \rho}^{2\,1}&-\Pi_{\omega \rho}^{2 \,2}&-\Pi_\rho^{2
\,1}&\textsf{p}^2-m_\rho^2-\Pi_\rho^{2\,2}\end{array} \right), \nonumber\\
\nonumber \end{eqnarray}
\begin{eqnarray} M_L&=& \left(
\begin{array}{cccc}
\textsf{p}^2-m_\pi^2-\Pi_\pi&-\Pi_{\pi \omega}&-\Pi_{\pi \rho}\\
\\
-\Pi_{\pi
\omega}&\textsf{p}^2-m_\omega^2-\Pi_\omega^{3\,3}&-\Pi_{\omega
\rho}\\ \\
-\Pi_{\pi \rho}&-\Pi_{\omega
\rho}&\textsf{p}^2-m_\rho^2-\Pi_\rho^{3\,3}
\end{array} \right)
\nonumber
\end{eqnarray}

For given values of the baryonic number density and the field
intensity, the equations Re$\left(\varepsilon_{S,T,L}\right)=0$
evaluated at ${\bm p}=0$ are equations in $p_0$, whose solutions are
identified as the effective masses of the mesons. For each equation
we have found multiple solutions, almost all the branches can be
traced back at zero density and identified with a definite meson
$\pi, \sigma, \omega$ or $\rho$.

In the next section we show the results obtained for two different
configurations of matter, symmetric nuclear matter ($n_1=n_2$) and
stellar matter composed of a neutral combination of electrons,
protons and neutrons in equilibrium against beta decay.

\section{Results and discussion}

In this section we analyze the effective meson masses for different
situations of physical interest. We consider magnetic intensities
$10^{16}-10^{19}$ G, a range that can be found in magnetars, and
matter at zero temperature and baryonic densities below $0.45$
fm$^{-3}$. Under these conditions strange baryons do not have a
significative role, therefore we only consider protons and neutrons.

For the model parameters we use the set Z271v6 of Ref.
\cite{HOROWITZ}, $m_\sigma=465$ MeV, $m_\omega=783$ MeV,
$m_\rho=763$ MeV, $g_\sigma=7.0313, g_\omega=8.406, g_\rho=10.016,
C_v=49.941, D=283.569, B=63.691$ and $A=1072.37$ MeV. This
parametrization guarantees the binding properties of nuclear
matter in the MFA ($n_0=0.1484$ fm$^{-3}$, $E_B=-16.24$ MeV,
$K=271$ MeV) and the viability of the direct URCA cooling in
neutron stars with mass 1.4 $M_\odot$.\\
We have added the pion-nucleon vertices, which do not modify the
MFA results.\\

As a first step we evaluate the MFA at zero temperature, in which
case the Fermi occupation number becomes a step function. As a
consequence, the Landau levels of the proton are occupied until a
well defined maximum value. At the end of this calculation we obtain
the chemical potentials, the effective nucleon mass, and the maximum
Landau level occupied as functions of the magnetic intensity and the
baryonic density.  The results of the MFA are inserted in the
neutron and proton propagators, for evaluating the polarization
insertions and the dielectric functions (\ref{DielS})-(\ref{DielL}).
We first investigate solutions corresponding to $p_0< 1$ GeV.\\
In Fig.2 the solutions as a function of the baryonic number
density for symmetric nuclear matter are shown for several
magnetic intensities. The neutral pion branch (a) increases with
the density, growing at most $10 - 15 \%$ for $n/n_0 = 3$. In Ref.
\cite{AGUIRRE} the author presented a similar calculation, but
neglecting the effect of heavier neutral mesons. A comparison of
Fig. 6c of that reference shows that for the same conditions a
more pronounced increase about $15-40 \%$ is obtained. Thus our
first conclusion is that the mix with the longitudinal vector
meson channels in the present calculations are responsible for a
strong moderation of the rate of growth with density at constant
magnetic intensity.\\
Returning to Fig.2a, it can be observed that the curves
 for intensities $10^{16} - 10^{18}$ G are very similar.
For a fixed density the slope decreases with $B$ until $B > 5
\times 10^{18}$ G where it increases rapidly. In fact, the highest
variation occurs between the curves corresponding to $5 \times
10^{18}$ G and $10^{19}$ G.

The $\sigma$ meson mass increases with density also, and a mean
growth of $15 \%$ is registered at $n/n_0=3$. But in contrast with
the previous case, these results scarcely depend on the field
intensity, showing at most $2 \%$ of dispersion with $B$ at the
highest density examined here.

The longitudinal component of the vector mesons also exhibit a
monotonous increasing trend with $n$. The growth of the $\rho$
meson mass reaches $20 \%$ while for the $\omega$ meson exceeds
$30 \%$. As can be clearly seen in Fig.2c, the dispersion due to
the magnetic field increases with the density, reaching $12 \%$ at
$n/n_0\simeq 1$ for the $\omega$ and $7 \%$ for the $\rho$ at
$n/n_0\simeq 2.8$. It is interesting that the curves corresponding
to $B=10^{19}$ G exhibits an almost vertical slope for certain
densities, and the disappearance of the $\rho$ branch after this
threshold.

A contrasting result is obtained for the transversal component of
the vector mesons (Fig.2d). The $\rho$ branch is almost constant
for all the range of densities and the dependence on $B$ is
negligible. The transversal $\omega$ branch, exhibits at
sub-saturation densities ( $n/n_0 < 0.5$ ) a behavior similar to
the corresponding longitudinal component. However, all the curves
change their concavity and become decreasing for high enough
densities.

In Fig.3 the same quantities as in Fig.2 are analyzed, but now for
a configuration of matter as can be found in a neutron star, that
is neutrons, protons and electrons in a homogeneous and neutral
compound in equilibrium against beta decay. \\
In the case of the pion mass (Fig.3a), the main differences with
respect to the previous case are {\it i}) the slope of the curves
increases with $B$ for $n/n_0> 0.75$, {\it ii}) in the present
case there is a stronger growth for $B > 5 \times 10^{18}$ G,
which causes an enhancement of $20 \%$ at $n/n_0=3$ and
$B=10^{19}$ G.

The $\sigma$ branch (Fig.3b) does not exhibit noticeable
differences, with the exception of the slight inversion in the
ordering of the curves for $B$ constant.

The longitudinal component of vector mesons (Fig.3c) presents
quantitative differences mainly in the $\rho$ branch. A weaker
growth with density, but a wider dispersion with $B$ at fixed
density correspond to the present case. The $\omega$ branch keeps
a strong growth, exceeding the $1$ GeV limit adopted for our
analysis at densities around $n/n_0=1.5$.

Finally for the transversal component of the vector mesons (Fig.
3d) we can see the $\rho$ branch remains practically invariable,
while the $\omega$ branch shows a slightly increased rate of
growth with density for $n/n_0 < 1.5$ but finally becomes
decreasing for higher densities.

In Figs. 4 and 5 we show the  dependence of the meson masses on
the magnetic intensity for isospin symmetric matter at a fixed
density $n/n_0=1$. In Fig.4 we include the results for the
intensities $B$ discussed up to this point, but in Fig.5 we extend
our analysis to larger values.\\
In Fig.4 we see that only the pion and the longitudinal $\omega$
component have a noticeable variation of roughly $3 \%$, while the
others remain almost invariable. Interestingly, the pion mass
decreases for low intensities,  until it reaches a stationary point
at $B \simeq 5 \times 10^{18}$ G and becomes increasing for stronger
fields.

 Up to this point we have examined the results valid for
physical interpretation. With the sole purpose of exploring the
limits of the approach used, in the following we expand this
panorama. In Fig.5 the range for the masses of resonances is broaden
up to $1.6$ GeV, and extreme magnetic intensities such that $e B
\sim (1.5 \Lambda_{QCD})^2$ are considered. Under such conditions
the lowest lying state is two-folded. It has a low $B$
manifestation, which we identify as the normal pion because it can
be continuously traced back at zero density to the empirical value
$m_\pi$ of the pion mass. It has just been described in Fig.4, but
here is evident that it ends abruptly at $e B \simeq 0.08$ GeV$^2$.
The second branch is identified as the abnormal pion although it is
strictly a resonance of the hadronic system. It starts near $e B
\simeq 0.05$ GeV$^2$ with an initial value $25 \%$ below the vacuum
value of the pion mass. Both branches coexist while they increase
rapidly. For greater intensities the abnormal branch stabilizes and
becomes the only pionic manifestation, with an effective mass
slightly below $m_\pi$.\\
The behavior of the neutral pion mass for a hadronic model with
pseudo-vector coupling in the presence of a magnetic field was
discussed in Ref. \cite{ADHYA}. Their results correspond to zero
baryonic density, so the comparison must be done carefully. In
fact it can be seen in Fig.2a, that magnetic effects are weaker as
the density approaches to zero. The results of \cite{ADHYA}
correspond to a monotonous and slight increase with $B$, showing a
growth below $1 \%$ at $e B = 0.02$ GeV$^2$. A similar trend can
be deduced from Fig.5b of  Ref. \cite{AGUIRRE} whose results
differ from the present calculations  in the inclusion of
additional meson fields. Therefore, we conclude that the meson
mixing changes considerably the dynamics of the neutral pion field
within the range $B < 10^{19}$ G and moderate baryonic densities.

Both components of the neutral $\rho$ meson as well as the
$\sigma$ meson masses  are really insensitive to the external
field showing a smooth, almost constant behavior. The transversal
$\omega$ branch is also linearly dependent on the field intensity
with a
weak slope. \\
The longitudinal branch of the omega meson collapses near $e B =
0.08$ GeV$^2$. Peculiarly, at this point it meets the branch
corresponding to a heavier partner resonance. The pair displays a
specular behavior, the meson mass increases with $B$ while the
mass of the partner has the opposite trend. Finally they coalesce
and disappear for a typical intensity.

The present work analyzes the effects on the effective meson
masses due to a combination of the external field and the response
of the dense hadronic medium. The contributions at zero density of
the fermion loops have not been considered. However, different
studies have focused on this subject, with different conclusions.
At zero density the neutral pion mass decreases with $B$ within
the NJL model \cite{NJL2}. In contrast the hadronic model of Ref.
\cite{ADHYA} predicts under the same conditions, a slight
increase. The same NJL treatment predicts an increase of the
$\sigma$ meson mass, whose magnitude is considerably greater than
for the pion.

\section{Conclusions}
In this work an analysis of the effective mass of the lightest
neutral mesons in the presence of an external magnetic field has
been carried out. The magnitude of the field as well as the
density of the hadronic medium have been taken as significative
variables, in a zero temperature treatment. Two different
configurations of the hadronic environment have been considered,
isospin symmetric nuclear matter and stellar matter. We have
focused on strong magnetic fields $10^{16}$ G $\leq B \leq
10^{19}$ G, and a range of baryonic densities where hyperons are
not expected to play a significative role.

The calculations have been made within a covariant model of
meson-baryon couplings plus meson self-interactions. The meson
polarization has been evaluated in a one loop self-consistent
calculation, but neglecting the divergent contributions from the
Dirac sea. The corrections due to the baryons have been introduced
by using a covariant propagator which includes the full effect of
the coupling to the external field through the electric charge and
the anomalous magnetic moments.

The analysis of the density dependence, shows that the composition
of the environment does not modify qualitatively the behavior of
the meson masses, and only minor features distinguish between
isospin symmetric nuclear matter and stellar matter.\\
The masses of all the mesons increase monotonically with the
density, with the exceptions of the transversal component of the
$\omega$ which becomes decreasing for high densities, and the
$\rho$ one, which is almost constant. The increase experienced at
the highest density is moderate (around $10 \%$) for the $\pi,
\sigma$ and transversal $\omega$, considerable for the
longitudinal $\rho$ (around $20 \%$), and important for the
longitudinal $\omega$, which exceeds the $30 \%$ in the mid-range
of densities.

The study of the dependence on the magnetic intensity for isospin
symmetric nuclear matter at the characteristic density $n=n_0$ in
the domain $10^{17}$ G $\leq B \leq 10^{19}$ G, and effective
meson masses below 1 GeV, shows a constant mass for the
transversal $\rho$, a sustained but scarce increase for the
transversal $\omega$ and the longitudinal $\rho$ branches. The
$\sigma$ and $\pi$ masses exhibit a non-monotonous and mirror-like
trend. For low intensities the $\pi$ mass decreases and the
$\sigma$ one increases with $B$, for medium to large intensities
the roles are changed. The largest variance of roughly $2 \%$
corresponds to the $\pi$ and longitudinal $\omega$.

Further developments for this line of investigation will include
the study of thermal effects and of the influence of the internal
structure of hadrons in the zero density state.

\section{Acknowledgements}
This work has been partially supported by a grant from the Consejo
Nacional de Investigaciones Cientificas y Tecnicas,  Argentina.

\section{Appendix A: One-loop meson polarization}\label{AppA}

Here the formulas for the nucleon loops contributing to the
different meson channels are shown, evaluated at zero spatial
component of the external momentum ${\bm p}=0$. For each channel
there is a sum of two contributions, due to neutron and proton
loops:

a) Neutron Contributions:

\begin{eqnarray}
%
\text{Re} \, \Pi_\pi (p_0)&=&\left(\frac{g_A}{4 \pi f_\pi}
\right)^2 \sum_s \int_0^{\infty} \, dt \, \Theta\left(
\tilde{\mu}_n-|{\cal M}_s|\right) \left( s s'+
\frac{m^2-t}{\Delta^2}\right) \Bigg\{ \left({\cal M}_{s'}^2-{\cal
M}_s^2\right)\log\left| \frac{\tilde{\mu}_p+p_{F n
s}}{\tilde{\mu}_p-p_{F n s}} \right| \nonumber\\
&+&\frac{p_0^2-\left({\cal M}_s-{\cal M}_{s'}\right)^2}{p_0
\lambda}\; \left( {\cal M}_s+{\cal M}_{s'}\right)^2 F(\alpha)\Bigg\}\nonumber \\
\text{Re} \, \Pi_\sigma (p_0)&=&\left(\frac{g_\sigma}{4
\pi}\right)^2 \sum_{s,s'} \int_0^{\infty} dt \, \Theta\left(
\tilde{\mu}_n-|{\cal M}_s|\right) \left( 1+s s'
\frac{m^2-t}{\Delta^2}\right)\Bigg\{ \log\left|
\frac{\tilde{\mu}_n+p_{F
s}}{\tilde{\mu}_n-p_{F s}} \right|\nonumber \\
&+&\frac{p_0^2-({\cal M}_s+{\cal M}_{s'})^2}{p_0 \lambda}
F(\alpha) \Bigg\}\nonumber \\
\text{Re} \, \Pi_\omega^{1 1} (p_0)&=&\left(\frac{g_\omega}{ 4
\pi}\right)^2 \sum_{s, s'} \int_0^{\infty} \, dt \, \Theta\left(
\tilde{\mu}_n-|{\cal M}_s|\right)\left(1-s' s \;
\frac{m^2}{\Delta^2}\right) \Bigg\{ \log\left|
\frac{\tilde{\mu}_n+p_{F
s}}{\tilde{\mu}_n-p_{F s}} \right|\nonumber \\
&+& s' s \;\frac{p_0^2-({\cal M}_s+{\cal M}_{s'})^2}{p_0
\lambda} F(\alpha)\Bigg\} \nonumber \\
\text{Re} \, \Pi_\omega^{1 2} (p_0)&=&-\frac{g_\omega}{8 \pi}
 m \sum_s s \int_0^{\infty} \frac{dt}{ \Delta} \frac{(2 \kappa_n B)^2-p_0^2}{p_0 \lambda}
\Bigg\{ \Theta\left(-p_0^2-2 p_0 |s \Delta+\kappa_n B|-4 s \Delta
\kappa_n B\right) \nonumber \\
&\times&\Theta\left(\tilde{\mu}_n+\frac{p_0^2+4 s \Delta \kappa_n
B}{2 p_0} \right)+\Theta\left(\tilde{\mu}_n-\frac{p_0^2-4 s \Delta
\kappa_n B}{2 p_0} \right) \left[1-2
\Theta\left(\tilde{\mu}_n+\frac{p_0^2+4 s \Delta \kappa_n B}{2
p_0} \right) \right] \nonumber \\
&\times& \Theta\left( p_0^2-2 p_0 |s
\Delta-\kappa_n B|-4 s \Delta \kappa_n B\right)\Bigg\} \nonumber \\
\text{Re} \, \Pi_\omega^{3 3} (p_0)&=&\left(\frac{g_\omega}{
\pi}\right)^2 \sum_{s,s'} \int_0^{\infty} \frac{dt}{2 p_0}
\delta_{s s'} \Theta\left( \tilde{\mu}_n-|{\cal M}_s|\right)
\frac{{\cal M}_s^2}{\lambda}F(\alpha) \nonumber
\end{eqnarray}
\begin{eqnarray}
\text{Re} \, \Pi_{\pi \omega} (p_0)&=&-\frac{g_\omega}{ \pi^2}
\frac{g_A}{f_\pi} m \sum_{s s'} s \, \delta_{s s'}  \int_0^{\infty}
\frac{dt}{2 \Delta} \Theta\left( \tilde{\mu}_n-|{\cal M}_s|\right)
\frac{{\cal
M}_s^2}{\lambda} F(\alpha) \nonumber \\
F(\alpha)&=&2 \Theta\left( 4 {\cal M}_s^2-(p_0-\alpha)^2 \right)\;
\arctan
\left(\frac{(p_0-\alpha) \,p_{F s}}{ \tilde{\mu}_n \lambda}\right)\nonumber\\
&+& \Theta\left( (p_0-\alpha)^2-4 {\cal M}_s^2 \right)\log
\left|\frac{ \tilde{\mu}_n \lambda-(p_0-\alpha) \,p_{F s}}{
\tilde{\mu}_n \lambda + (p_0-\alpha) \,p_{F s}} \right| \nonumber
\end{eqnarray}

where $\Delta=\sqrt{m^2+t}$, ${\cal M}_s=s \Delta-\kappa_n B$,
$p_{Fs}=\sqrt{\tilde{\mu}_n^2-{\cal M}_s^2}$, $\alpha=({\cal
M}_{s'}^2-{\cal M}_s^2)/p_0$, and $\lambda=\sqrt{|4 {\cal
M}_s^2-(p_0-\alpha)^2|}$.
  Although the
domain of integration is not bounded, the relation
\[\Theta\left(\tilde{\mu}_n-|{\cal M}_s|\right)\equiv
\Theta\left((\tilde{\mu}_n+s \kappa_n B)^2-m^2-t\right)\,
\Theta\left(\tilde{\mu}_n+s \kappa_n B-m\right) \] valid for the
conditions under consideration, establishes an upper limit of
integration.\\

b) Proton contributions

\begin{eqnarray}
\text{Re} \, \Pi_\sigma (p_0)&=&\left(\frac{g_\sigma}{4
\pi}\right)^2 q B \sum_{s,s';l,n=0} \delta_{n l}\;
\Theta\left(\tilde{\mu}_p-|{\cal M}_{n s}|\right)  \left(1+2\,
\frac{m-s'\Delta_n}{m+s\Delta_n}+\frac{m-s\Delta_n}{m+s\Delta_n}\frac{m-s'\Delta_n}{m+s'\Delta_n}\right)\nonumber \\
&\times&\frac{\Delta_n+s \,m}{\Delta_n}\frac{\Delta_n+s'
m}{\Delta_n}
 \Bigg\{\log\left| \frac{\tilde{\mu}_n+p_{F s}}{\tilde{\mu}_n-p_{F s}}
\right|+\frac{p_0^2-\left({\cal M}_{n s}+{\cal M}_{n
s'}\right)^2}{\eta
p_0 \lambda_{l s', n s}}\; G_{l s', n s}\Bigg\}\nonumber \\
\text{Re} \, \Pi_\pi (p_0)&=&\left(\frac{g_A}{4 \pi f_\pi}
\right)^2 q B \sum_{s,s';l,n=0} \delta_{n l}\;
\Theta\left(\tilde{\mu}_p-|{\cal M}_{n s}|\right)\left(1-2\,
\frac{m-s'\Delta_n}{m+s\Delta_n}+\frac{m-s\Delta_n}{m+s\Delta_n}\frac{m-s'\Delta_n}{m+s'\Delta_n}\right)\nonumber \\
&\times& \frac{\Delta_n+s \,m}{\Delta_n}\frac{\Delta_n+s'
m}{\Delta_n}\Bigg\{ \left({\cal M}_{n s}^2-{\cal M}_{n
s'}^2\right)\log\left(\frac{\tilde{\mu}_p+p_{F n
s}}{\tilde{\mu}_p-p_{F n s}} \right)+\frac{p_0-\alpha_{l s', n
s}}{\lambda_{l s', n s}}
\left( {\cal M}_{n s}+{\cal M}_{n s'}\right)^2 \nonumber \\
&\times& \; G_{l s', n s}\Bigg\}\nonumber \\
\text{Re} \, \Pi_\omega^{1 1} (p_0)&=&\left(\frac{g_\omega}{ 4
\pi}\right)^2 q B \sum_{s,s';l,n=1}\left(\delta_{n+1,l}\,
\frac{\Delta_n+s m}{\Delta_n}\;\frac{\Delta_l- s'
m}{\Delta_l}+\delta_{n,l+1}\, \frac{\Delta_n-s
m}{\Delta_n}\;\frac{\Delta_l+ s m}{\Delta_l}\right)\nonumber \\
&\times& \Theta\left( \tilde{\mu}_p-|{\cal M}_{n s}|\right) \Bigg[
\log\left| \frac{\tilde{\mu}_p+p_{F n s}}{\tilde{\mu}_p-p_{F n
s}}\right| +\frac{p_0^2-\left({\cal M}_{n s}+{\cal M}_{l
s'}\right)^2}{p_0 \lambda_{l s', n s}}\; G_{n s, l s'}\Bigg]
\nonumber \\
\text{Re} \, \Pi_\omega^{1 2} (p_0)&=&-\frac{g_\omega^2}{ 16 \pi}\,
q B \sum_{s,s',l,n=1} \left(\delta_{n+1,l} \, \frac{\Delta_n+s
m}{\Delta_n}\;\frac{\Delta_l- s'm}{\Delta_l}- \delta_{n,l+1} \,
\frac{\Delta_n-s m}{\Delta_n}\;\frac{\Delta_l+ s'm}{\Delta_l}\right)
\nonumber \\
\times&\Bigg\{& \Theta\left( \tilde{\mu}_p+\frac{p_0-\alpha_{n s, l
s'}}{2}\right) \Theta\left( -\frac{p_0-\alpha_{n s, l s'}}{2}-{\cal
M}_{l s'}\right) + \Theta\left( \tilde{\mu}_p-\frac{p_0+\alpha_{n s,
l s'}}{2}\right) \nonumber \\
&\times& \Theta\left( \frac{p_0+\alpha_{n s,l s'}}{2}-{\cal M}_{n
s}\right)\left[1-2 \Theta\left( \tilde{\mu}_p+\frac{p_0-\alpha_{n s,
l s'}}{2}\right) \right]\Bigg\} \frac{p_0^2-\left({\cal M}_{n
s}+{\cal
M}_{l s'}\right)^2}{p_0 \lambda_{l s', n s}}\nonumber \\
\text{Re} \, \Pi_\omega^{3 3} (p_0)&=&\left(\frac{g_\omega}{
\pi}\right)^2 \frac{q B}{p_0} \sum_{s,s';l,n=0} \delta_{n l}\;
\delta_{s s'} \; \Theta\left( \tilde{\mu}_p-|{\cal M}_{n s}|\right)
\frac{{\cal M}_{n s}^2}{\lambda_{l s', n s}} \; G_{l s', n s}
\nonumber
\end{eqnarray}
\begin{eqnarray}
\text{Re} \, \Pi_{\pi \omega} (p_0)&=&\frac{g_\omega}{ \pi^2}
\frac{g_A}{f_\pi}\, q B m \sum_{s s',n=0} s \, \delta_{s s'}
\delta_{n l} \Theta\left( \tilde{\mu}_p-|{\cal M}_{n s}|\right)
\frac{{\cal M}_{n s}^2}{\Delta_n \lambda_{l s', n s}} G_{l s', n s}
\nonumber
\end{eqnarray}
\begin{eqnarray}
G_{l s', n s}&=&2\, \Theta\left(4 {\cal
 M}_{l s'}^2-(p_0+\alpha_{l s', n s})^2\right)
 \arctan\left( p_{F l s'}\frac{p_0+\alpha_{l s', n s}}{\tilde{\mu}_p
 \lambda_{l s', n s}}\right)\nonumber \\
&+& \Theta\left((p_0+\alpha_{l s', n s})^2-4 {\cal
 M}_{l s'}^2\right)\log\left|\frac{\tilde{\mu}_p \lambda_{l s', n s}- (p_0+\alpha_{l s', n s})p_{F l s'}}
 {\tilde{\mu}_p \lambda_{l s', n s}+ (p_0+\alpha_{l s', n s})p_{F l s'}} \right|\nonumber
\end{eqnarray}

with $\Delta_n=\sqrt{m^2+2 n q B}$, ${\cal M}_{n s}=s
\Delta_n-\kappa_p B$, $p_{F n s}=\sqrt{\tilde{\mu}_p^2-{\cal M}_{n
s}^2}$, $\alpha_{l s', n s}=({\cal M}_{l s'}^2-{\cal M}_{n
s}^2)/p_0$, and $\lambda_{l s', n s}=\sqrt{|4 {\cal M}_{l
s'}^2-(p_0+\alpha_{l s', n s})^2|}$ . In the summations it must be
understood that for $n=0$ only the case $s=s'=1$ must be included.

Furthermore, the following relations are valid for both protons
and neutrons
\begin{eqnarray}
\text{Re} \, \Pi_\omega^{2 2} (p_0)&=&\text{Re} \, \Pi_\omega^{1 1} (p_0), \nonumber \\
\text{Re} \, \Pi_\rho^{a b} (p_0)&=&\left(\frac{g_\rho}{ 2
g_\omega}\right)^2\text{Re} \, \Pi_\omega^{a b} (p_0), \;\;
a,b=1,2,3\nonumber
\end{eqnarray}
whereas there is a change of sign for
\begin{eqnarray}
\text{Re} \, \Pi_{\omega \rho}^{a b} (p_0)&=&\pm \frac{g_\rho}{2 g_\omega} \text{Re} \, \Pi_\omega^{a b} (p_0), \; a,b=1,2,3 \nonumber \\
\text{Re} \, \Pi_{\pi \rho} (p_0)&=&\pm\frac{g_\rho}{ 2
g_\omega}\text{Re} \, \Pi_{\pi \omega} (p_0)\nonumber
\end{eqnarray}
corresponding the upper (lower) sign to protons (neutrons).

The effective chemical potential $\tilde{\mu}$ is related with the
thermodynamical chemical potential by means of
$\tilde{\mu}_{n,p}=\mu_{n,p}-g_\omega W \pm g_r R/2$.

\newpage
\begin{figure}
\includegraphics[width=0.8\textwidth]{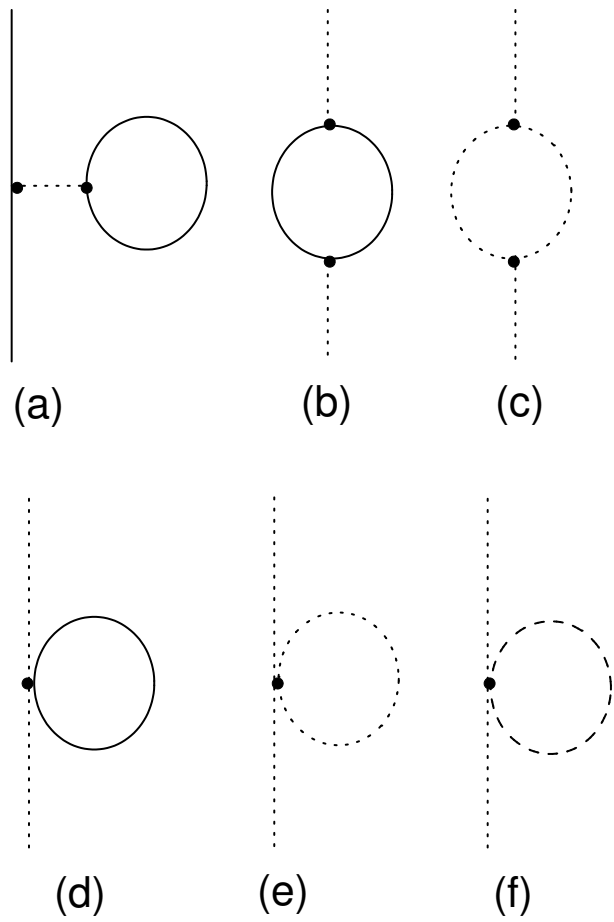}
\caption{\footnotesize The diagrams considered in our
Dyson-Schwinger calculations. Solid lines represent fermion
propagators, and dashed or dotted lines the meson ones. }
\end{figure}

\newpage
\begin{figure}
\includegraphics[width=0.8\textwidth]{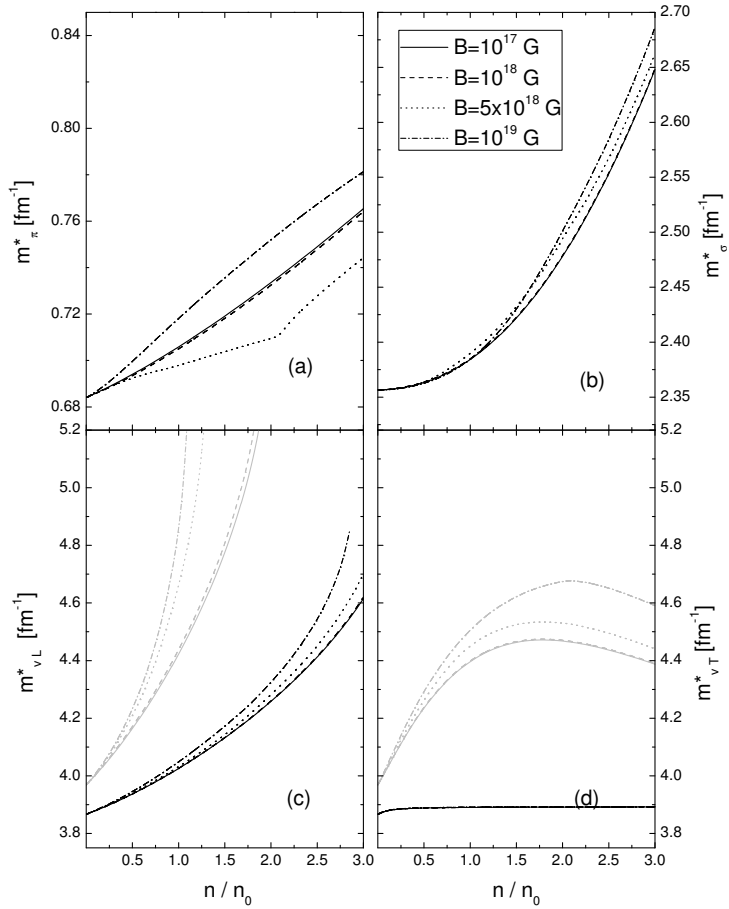}
\caption{\footnotesize The effective meson masses as functions of
the baryonic number density for several magnetic intensities
corresponding to isospin symmetric nuclear matter. }
\end{figure}

\newpage
\begin{figure}
\includegraphics[width=0.8\textwidth]{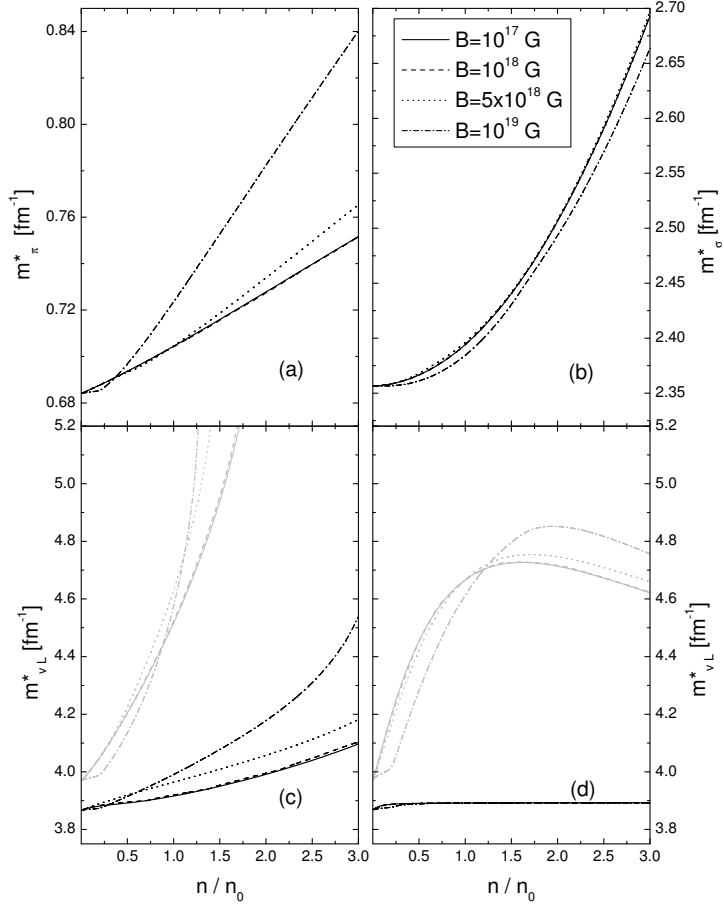}
\caption{\footnotesize The effective meson masses as functions of
the baryonic number density for several magnetic intensities
corresponding to neutral stellar matter. }
\end{figure}

\newpage
\begin{figure}
\includegraphics[width=0.8\textwidth]{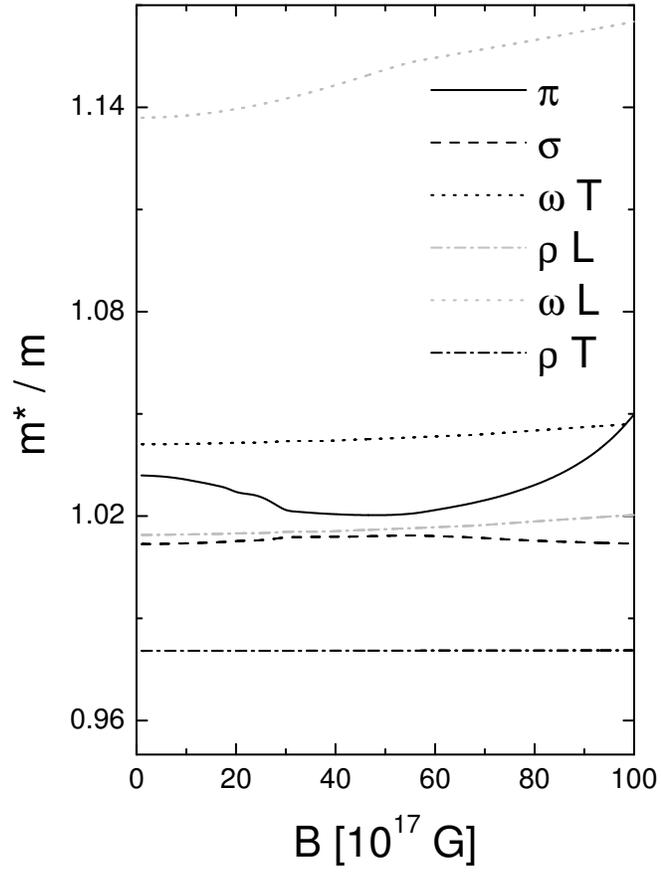}
\caption{\footnotesize The effective meson masses as functions of
the  magnetic intensity for isospin symmetric nuclear matter at
the fixed baryonic number density $n=n_0$. }
\end{figure}

\newpage
\begin{figure}
\includegraphics[width=0.8\textwidth]{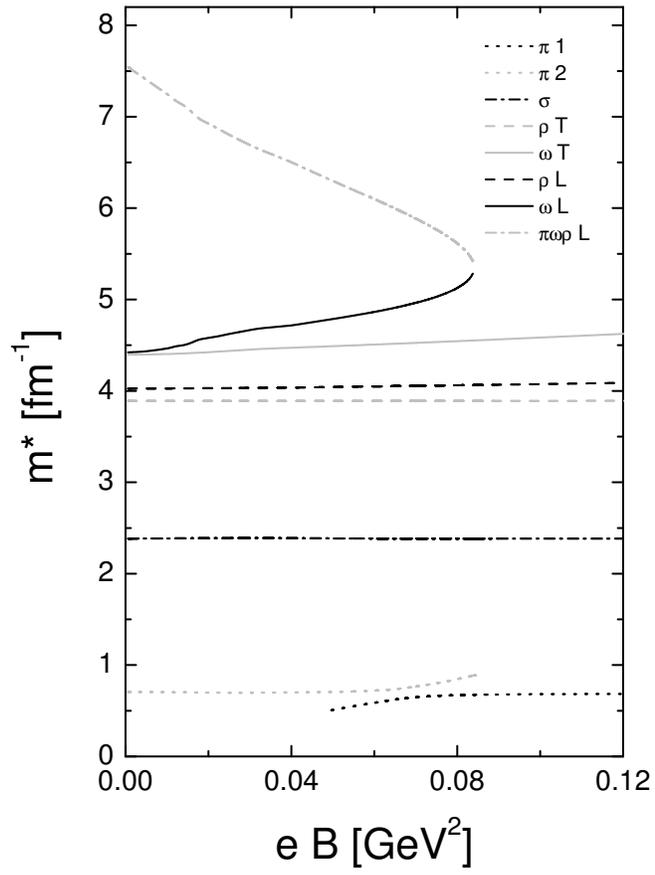}
\caption{\footnotesize The same as in Fig. 4, but for a wider
magnetic range and an extended scale for the meson masses. }
\end{figure}

\end{document}